\documentclass[twocolumn,A4]{article}
\usepackage[dvips]{graphics}
\begin{document}
\onecolumn
\begin{center}
{\bf{\Large Electron transport through honeycomb lattice ribbons
with armchair edges}}\\
~\\
Santanu K. Maiti$^{1,2,*}$ \\
~\\
{\em $^1$Theoretical Condensed Matter Physics Division,
Saha Institute of Nuclear Physics, \\
1/AF, Bidhannagar, Kolkata-700 064, India \\
$^2$Department of Physics, Narasinha Dutt College,
129, Belilious Road, Howrah-711 101, India} \\
~\\
{\bf Abstract}
\end{center}
We address electron transport in honeycomb lattice ribbons with armchair
edges attached to two semi-infinite one-dimensional metallic electrodes 
within the tight-binding framework. Here we present numerically the 
conductance-energy and current-voltage characteristics as functions of 
the length and width of the ribbons. Our theoretical results predict that 
for a ribbon with much smaller length and width, so-called a nanoribbon, 
a gap in the conductance spectrum appears across the energy $E=0$. While, 
this gap decreases gradually with the increase of the size of the ribbon, 
and eventually it almost vanishes. This reveals a transformation from the 
semiconducting to the conducting material, and it becomes much more clearly 
visible from our presented current-voltage characteristics.

\vskip 1cm
\begin{flushleft}
{\bf PACS No.}: 73.63.-b; 73.63.Rt.  \\
~\\
{\bf Keywords}: Honeycomb lattice ribbon; Armchair edges; Conductance; 
$I$-$V$ characteristic.
\end{flushleft}
\vskip 4in
\noindent
{\bf ~$^*$Corresponding Author}: Santanu K. Maiti

Electronic mail: santanu.maiti@saha.ac.in
\newpage
\twocolumn

\section{Introduction}

The electronic transport in nanoribbons of graphene has opened up new areas
in nanoelectronics. A graphene nanoribbon (GNR) is a monolayer of carbon 
atoms arranged in a honeycomb lattice structure~\cite{falko,pen,geim,berg}. 
Due to the special electronic and physical properties, graphene based 
materials exhibit several novel properties like unconventional quantum Hall 
effect~\cite{zhang}, high carrier mobility~\cite{geim} and many others. The 
high carrier mobility in graphene demonstrates the idea for fabrication of 
high speed switching devices those have widespread applications in different 
fields. Some recent experiments~\cite{ouy,yan,are} have also suggested that 
GNRs can be used to design field-effect transistors and this application 
provides a huge interest in the community of nanoelectronics device research. 
Furthermore, GNRs can be used to construct MOSFETs which perform much better 
than conventional Si MOSFETs. In other experiment~\cite{li} it has been 
proposed that a narrow strip of graphene with armchair edges, so-called a 
graphene nanoribbon, exhibits semiconducting behavior due to its edge 
effects, unlike carbon nanotubes of larger sizes which are mixtures of 
both metallic and semiconducting materials. This is due to the fact that 
in a narrow graphene sheet, a band gap appears across the energy $E=0$, 
while the gap gradually disappears with the increase of the size of the 
ribbon. This reveals a transformation from the semiconducting to the 
metallic material, and such a phenomenon can be utilized for fabrication 
of electronic devices. This motivates us to study the electron transport 
in honeycomb lattice ribbons with armchair edges and to verify qualitatively 
how the transformation from the semiconducting to the conducting property 
can be achieved simply by tuning the size of a ribbon.

The purpose of the present paper is to provide a qualitative study of 
electron transport in honeycomb lattice ribbons with armchair edges 
attached to two semi-infinite one-dimensional metallic electrodes (see 
Fig.~\ref{armchair}). The theoretical description of electron transport 
in a bridge system has been followed based on the pioneering work of Aviram 
and Ratner~\cite{aviram}. Later, many excellent experiments~\cite{tali,
reed1,reed2,novo1,novo2} have been done in several bridge systems to 
understand the basic mechanisms underlying the electron transport. Though 
in literature many theoretical~\cite{kat1,kat2,kat3,kat4,lian,hod,cre,cua,
cron,holl,orella1,orella2,nitzan1,nitzan2,new,muj1,muj2} as well as 
experimental papers~\cite{tali,reed1,reed2,novo1,novo2} on electron transport 
are available, yet lot of controversies are still present between the theory 
and experiment, and the complete knowledge of the conduction mechanism in 
this scale is not very well established even today. 
Several controlling parameters are there which can regulate significantly 
the electron transport in a conducting bridge, and all these effects have 
to be taken into account properly to reveal the transport properties. For our
illustrative purposes, here we describe very briefly some of these effects. \\
\noindent
(i) The quantum interference effect~\cite{baer1,baer2,baer3,tagami,walc1}
of electron waves passing through different arms of any conducting element 
which bridges two electrodes becomes the most significant issue. \\
\noindent
(ii) The coupling of 
the electrodes with bridging material provides an important signature in 
the determination of current amplitude across any bridge system~\cite{baer1}. 
The understanding of this coupling to the electrodes under non-equilibrium 
condition is a major challenge, and we should take care about it in 
fabrication of any electronic device. \\
\noindent
(iii) The geometry 
of the conducting material between the two electrodes itself is an important 
issue to control the electron transmission. To emphasize it, Ernzerhof 
{\em et al.}~\cite{ern2} have predicted several model calculations and 
provided some significant results. \\
\noindent
(iv) The dynamical fluctuation in 
the small-scale devices is another important factor which plays an active 
role and can be manifested through the measurement of {\em shot 
noise}~\cite{blanter,walc2}, a direct consequence of the quantization of 
charge. It can be used to obtain information on a system which is not 
available directly through the conductance measurements, and is generally 
more sensitive to the effects of electron-electron correlations than the 
average conductance.

Furthermore, several other parameters of the Hamiltonian that describe a 
system also provide significant effects in the determination of the current 
across a bridge system.

Here we adopt a simple tight-binding model to describe the system and all 
the calculations are performed numerically. We address the conductance-energy 
and current-voltage characteristics as functions of lengths and widths of
ribbons. Our results clearly predicts how a honeycomb lattice ribbon with
armchair edges transforms its behavior from the semiconducting to the 
metallic nature, and this feature may be utilized in fabrication of 
nanoelectronic devices.

The paper is organized as follow. Following the introduction 
(Section $1$), in Section $2$, we present the model and the theoretical 
formulations for our calculations. Section $3$ discusses the significant 
results, and finally, we summarize our results in Section $4$.

\section{Model and the synopsis of the theoretical background}

\begin{figure}[ht]
{\centering \resizebox*{7.5cm}{3.5cm}{\includegraphics{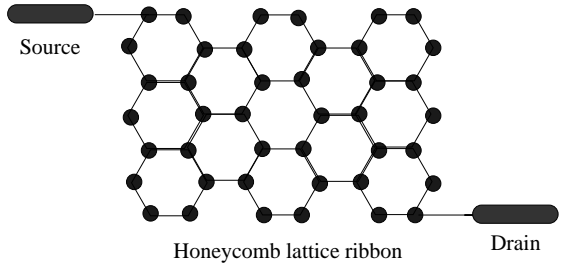}}\par}
\caption{Schematic view of a honeycomb lattice ribbon with armchair edges
attached to two semi-infinite one-dimensional metallic electrodes, viz, 
source and drain. Filled circles correspond to the position of the atomic
sites (for color illustration, see the web version).}
\label{armchair}
\end{figure}

Let us refer to Fig.~\ref{armchair}, where a honeycomb lattice ribbon with
armchair edges is attached to two semi-infinite one-dimensional metallic 
electrodes, viz, source and drain. It is important to note that throughout
this study we attach the electrodes at the two extreme ends of nanoribbons,
as seen in Fig.~\ref{armchair}, to keep the uniformity of the quantum 
interference effects.

To calculate the conductance $g$ of the ribbon, we use the Landauer
conductance formula~\cite{datta,marc}, and at very low temperature and
bias voltage it can be expressed in the form,
\begin{equation}
g=\frac{2e^2}{h} T
\label{equ1}
\end{equation}
where $T$ gives the transmission probability of an electron in the ribbon. 
This $(T)$ can be represented in terms of the Green's function of the 
ribbon and its coupling to the two electrodes by the 
relation~\cite{datta,marc},
\begin{equation}
T=Tr\left[\Gamma_S G_{rib}^r \Gamma_D G_{rib}^a\right]
\label{equ2}
\end{equation}
where $G_{rib}^r$ and $G_{rib}^a$ are respectively the retarded and advanced
Green's functions of the ribbon including the effects of the electrodes.
The parameters $\Gamma_S$ and $\Gamma_D$ describe the coupling of the
ribbon to the source and drain respectively, and they can be defined in
terms of their self-energies. For the full system i.e., the ribbon, source
and drain, the Green's function is defined as,
\begin{equation}
G=\left(\epsilon-H\right)^{-1}
\label{equ3}
\end{equation}
where $\epsilon=E+i\delta$. $E$ is the injecting energy of the source electron
and $\delta$ gives an infinitesimal imaginary part to $\epsilon$. To Evaluate
this Green's function, the inversion of an infinite matrix is needed since
the full system consists of the finite ribbon and the two semi-infinite 
electrodes. However, the entire system can be partitioned into sub-matrices 
corresponding to the individual sub-systems and the Green's function for 
the ribbon can be effectively written as,
\begin{equation}
G_{rib}=\left(\epsilon-H_{rib}-\Sigma_S-\Sigma_D\right)^{-1}
\label{equ4}
\end{equation}
where $H_{rib}$ is the Hamiltonian of the ribbon which can be written in 
the tight-binding model within the non-interacting picture like,
\begin{eqnarray}
H_{rib}=\sum_i \epsilon_i c_i^{\dagger} c_i + \sum_{<ij>} t 
\left(c_i^{\dagger} c_j + c_j^{\dagger} c_i\right)
\label{equ5}
\end{eqnarray}
In the above Hamiltonian ($H_{rib}$), $\epsilon_i$'s are the site energies, 
$c_i^{\dagger}$ ($c_i$) is the creation (annihilation) operator of an 
electron at the site $i$ and $t$ is the nearest-neighbor hopping integral.
Similar kind of tight-binding Hamiltonian is also used to describe the
two semi-infinite
one-dimensional perfect electrodes where the Hamiltonian is parametrized
by constant on-site potential $\epsilon_0$ and nearest-neighbor hopping
integral $t_0$. In Eq.~\ref{equ4}, $\Sigma_S=h_{S-rib}^{\dagger}g_S 
h_{S-rib}$ and $\Sigma_D=h_{D-rib} g_D h_{D-rib}^{\dagger}$ are the 
self-energy operators due to the two electrodes, where $g_S$ and $g_D$ 
correspond to the Green's functions of the source and drain respectively. 
$h_{S-rib}$ and $h_{D-rib}$ are the coupling matrices and they will be 
non-zero only for the adjacent points of the ribbon, and the electrodes 
respectively. The matrices
$\Gamma_S$ and $\Gamma_D$ can be calculated through the expression,
\begin{equation}
\Gamma_{S(D)}=i\left[\Sigma_{S(D)}^r-\Sigma_{S(D)}^a\right]
\label{equ6}
\end{equation}
where $\Sigma_{S(D)}^r$ and $\Sigma_{S(D)}^a$ are the retarded and advanced
self-energies respectively, and they are conjugate with each other.
These self-energies can be written as~\cite{tian},
\begin{equation}
\Sigma_{S(D)}^r=\Lambda_{S(D)}-i \Delta_{S(D)}
\label{equ7}
\end{equation}
where $\Lambda_{S(D)}$ are the real parts of the self-energies which
correspond to the shift of the energy eigenvalues of the ribbon and
the imaginary parts $\Delta_{S(D)}$ of the self-energies represent the
broadening of these energy levels. Since this broadening is much larger 
than the thermal broadening, we restrict our all calculations only
at absolute zero temperature. All the informations about the 
ribbon-to-electrode coupling are included into these two self-energies.

The current passing across the ribbon can be depicted as a single-electron
scattering process between the two reservoirs of charge carriers. The
current $I$ can be computed as a function of the applied bias voltage $V$
through the relation~\cite{datta},
\begin{equation}
I(V)=\frac{e}{\pi \hbar}\int_{E_F-eV/2}^{E_F+eV/2} T(E,V) dE
\label{equ8}
\end{equation}
where $E_F$ is the equilibrium Fermi energy. Here we make a realistic
assumption that the entire voltage is dropped across the ribbon-electrode
interfaces, and it is examined that under such an assumption the $I$-$V$
characteristics do not change their qualitative features. This assumption
is based on the fact that, the electric field inside the ribbon especially 
for narrow ribbons seems to have a minimal effect on the conductance-voltage 
characteristics. On the other hand, for quite larger ribbons and high bias 
voltages the electric field inside the ribbon may play a more significant 
role depending on the internal structure and size of the ribbon~\cite{tian}, 
but the effect becomes too small.

\section{Results and discussion}

In order to understand the dependence of electron transport on the lengths
and widths of nanoribbons, in the present article, we concentrate only on
the cleaned systems rather than any dirty one. Accordingly, we set the site 
energies of the honeycomb lattice ribbons 
as $\epsilon_i=0$ for all $i$. The values of the other parameters are
assigned as follow: the nearest-neighbor hopping integral $t$ in the ribbon
is set to $2$, the on-site energy $\epsilon_0$ and the hopping integral
$t_0$ for the two electrodes are fixed to $0$ and $2$ respectively. The
parameters $\tau_S$ and $\tau_D$ are set as $1.5$, where they correspond 
to the hopping strengths of the ribbon to the source and drain respectively.
In addition to these, we also introduce two other parameters $N$ and $M$
to reveal the size of a nanoribbon, where they correspond to the width 
and length of the ribbon respectively. Thus, for example, a nanoribbon 
with $N=1$ and $M=4$ represents a linear chain of four hexagons. Hence the
parameter $M$ determines the total number of hexagons in a single chain.
Following this rule, a nanoribbon with $N=3$ and $M=3$ corresponds to 
three linear chains attached side by side (see Fig.~\ref{armchair}) where 
each chain contains three hexagons. For simplicity, throughout our study 
we set the Fermi energy $E_F=0$ and choose the units where $c=e=h=1$.

\begin{figure}[ht]
{\centering \resizebox*{7.75cm}{10cm}{\includegraphics{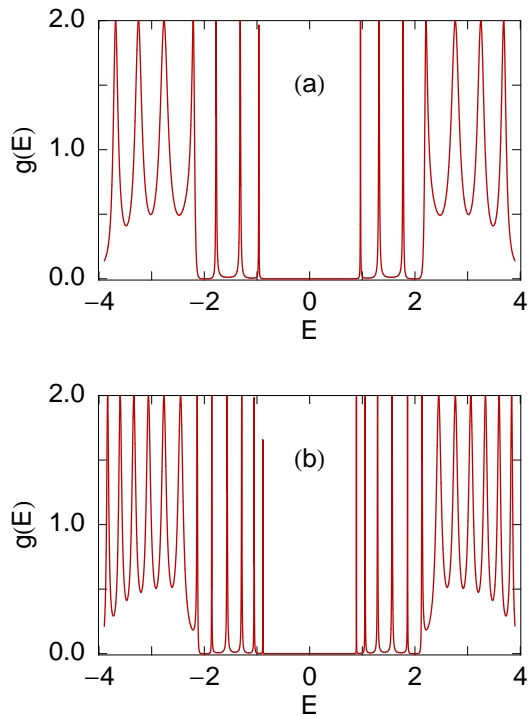}}\par}
\caption{Conductance $g$ as a function of the energy $E$ for some lattice
ribbons with fixed width $N=1$ and varying lengths where (a) $M=6$ and
(b) $M=10$ (for color illustration, see the web version).}
\label{cond1}
\end{figure}

\begin{figure}[ht]
{\centering \resizebox*{7.75cm}{10cm}{\includegraphics{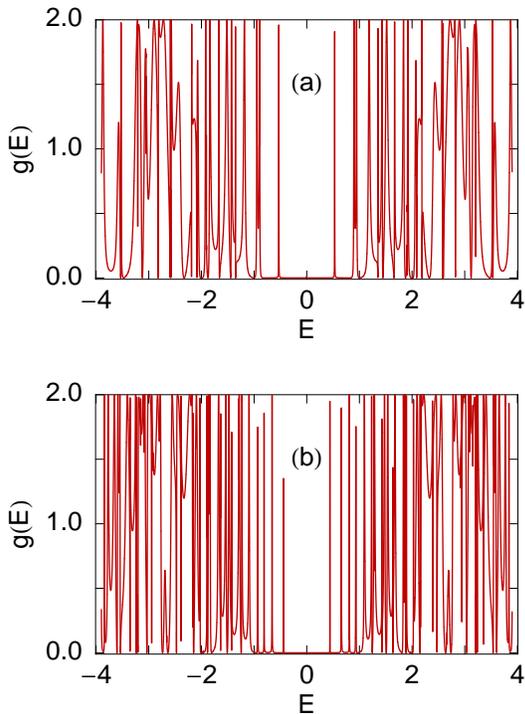}}\par}
\caption{Conductance $g$ as a function of the energy $E$ for some lattice
ribbons with fixed width $N=4$ and varying lengths (identical as in
Fig.~\ref{cond1}) where (a) $M=6$ and (b) $M=10$. Here the width of the 
ribbons is increased compared to the ribbons as taken in Fig.~\ref{cond1} 
(for color illustration, see the web version).}
\label{cond2}
\end{figure}

\begin{figure}[ht]
{\centering \resizebox*{7.75cm}{5cm}{\includegraphics{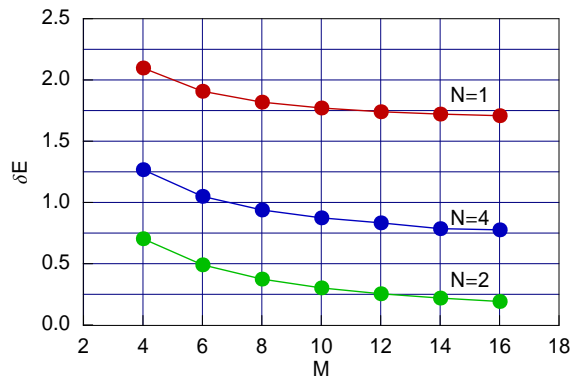}}\par}
\caption{Variation of the central energy gap $\delta E$ as a function of 
the length $M$ for some lattice ribbons with fixed widths $N$. The red,
green and blue curves correspond to $N=1$, $2$ and $4$ respectively
(for color illustration, see the web version).}
\label{gap}
\end{figure}

Let us first describe the variation of the conductance $g$ as a function of 
the injecting electron energy $E$. In Fig.~\ref{cond1} we present the
conductance-energy ($g$-$E$) characteristics for some honeycomb lattice
ribbons with fixed width ($N=1$) and varying lengths, where (a) and (b)
correspond to the linear chains with six ($M=6$) and ten ($M=10$) hexagons 
respectively. The conductance spectra shows fine resonance peaks for
some particular energies, while for all other values of the energy $E$,
either it ($g$) drops to zero or gets much small value. At these resonance
energies, the conductance gets the value $2$, and hence, the transmission 
probability $T$ becomes unity since the expression $g=2T$ holds from the
Landauer conductance formula (see Eq.~\ref{equ1}). These resonance peaks
are associated with the energy levels of the nanoribbons and thus the
conductance spectra, on the other hand, reveal the signature of the 
energy spectra of the
nanoribbons. The most important issue observed from these spectra is that,
a central gap appears across the energy $E=0$ and the width of the gap
becomes small for the chain with $10$ hexagons compared to the other
chain i.e., the chain with $6$ hexagons. It predicts that, for a fixed 
width, the central energy gap decreases with the increase of the length
of the nanoribbon. In the same footing, to visualize the
dependence of the width on the conductance-energy characteristics, in
Fig.~\ref{cond2} we display the results for some honeycomb lattice ribbons
considering the width $N=4$, where (a) and (b) represent the nanoribbons
with identical lengths as in Fig.~\ref{cond1}. The results show that,
due to the large system sizes the $g$-$E$ characteristics exhibit almost
a quasi-continuous variation across the energy $E=0$. For both these two
ribbons the energy gap also appears around the energy $E=0$, and the gap 
decreases with the increase of the length of the nanoribbon. Comparing the
results presented in Figs.~\ref{cond1} and \ref{cond2}, we can emphasize
that for a fixed width the central energy gap always decreases with the
size of the nanoribbon. Now to reveal the dependence of the energy gap on 
the system size much more clearly, in Fig.~\ref{gap} we
show the variation of the central energy gap $\delta E$ as a function of 
the length $M$ for some honeycomb lattice ribbons with different widths $N$. 
The red, green and blue lines correspond to the results for the ribbons
with fixed widths $N=1$, $2$ and $4$ respectively. These results clearly
emphasize that for the fixed width the gap gradually decreases with the
increase of the length of the nanoribbon. It is also examined that for
much larger lengths it ($\delta E$) almost vanishes (not shown here in
the figure). Quite similar nature is also observed if we plot the 
variation of the energy gap as a function of the length $N$ keeping the
width $M$ as a constant, and due to the obvious reason we do not plot these
results further in the present description. These results provide us an
important signature which concern with a transition from the semiconducting
(finite energy gap) to the conducting (zero energy gap) material, and this 
transition can be achieved simply by tuning the size of the nanoribbon. 

\begin{figure}[ht]
{\centering \resizebox*{7.75cm}{10cm}{\includegraphics{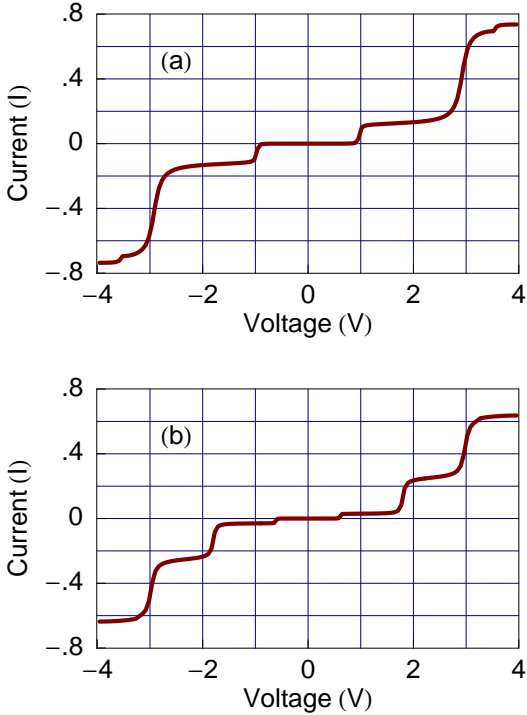}}\par}
\caption{Current $I$ as a function of the bias voltage $V$ for some lattice
ribbons with fixed width $N=2$ and varying lengths where (a) $M=3$ and
(b) $M=5$ (for color illustration, see the web version).}
\label{current1}
\end{figure}

All these basic features of electron transfer can be quite easily explained 
from our study of the current-voltage ($I$-$V$) characteristics rather than 
the conductance-energy spectra. The current $I$ is determined from 
the integration procedure of the transmission function ($T$) (see 
Eq.~\ref{equ8}), where the function $T$ varies exactly similar to the 
conductance spectra, differ only in magnitude by a factor $2$, since the 
relation $g=2T$ holds from the Landauer conductance formula (Eq.~\ref{equ1}).
The variation of the current-voltage characteristics for some typical
honeycomb lattice ribbons with fixed width $N=2$ and varying lengths 
is presented in Fig.~\ref{current1}, where (a) and (b) correspond to the 
ribbons with $M=3$ and $5$ respectively. The current exhibits a staircase
like behavior as a function of the applied bias voltage $V$. This staircase
like nature appears due to the existence of the resonance peaks in the 
conductance spectra since the current is computed by the integration
process of the transmission function $T$. As we increase the bias voltage
$V$, the electrochemical potentials in the two electrodes cross one of
the energy levels of the ribbon and accordingly a jump in the $I$-$V$
curve appears. The sharpness of the steps in the current-voltage 
characteristics and the current amplitude solely depend on the coupling 
strengths of the nanoribbon to the electrodes, viz, source and drain. It
is observed that, in the limit of weak coupling, defined by the condition 
$\tau_{S(D)} << t$, current shows staircase like structure with sharp 
steps. While, in the strong coupling limit,
described by the condition $\tau_{S(D)} \sim t$, current varies quite
continuously with the bias voltage $V$ and achieves large current amplitude
compared to the weak-coupling limit. All these coupling effects have 
clearly been explained in many papers in the literature. The significant
feature observed from the figure (Fig.~\ref{current1}) is that for the 
\begin{figure}[ht]
{\centering \resizebox*{7.75cm}{10cm}{\includegraphics{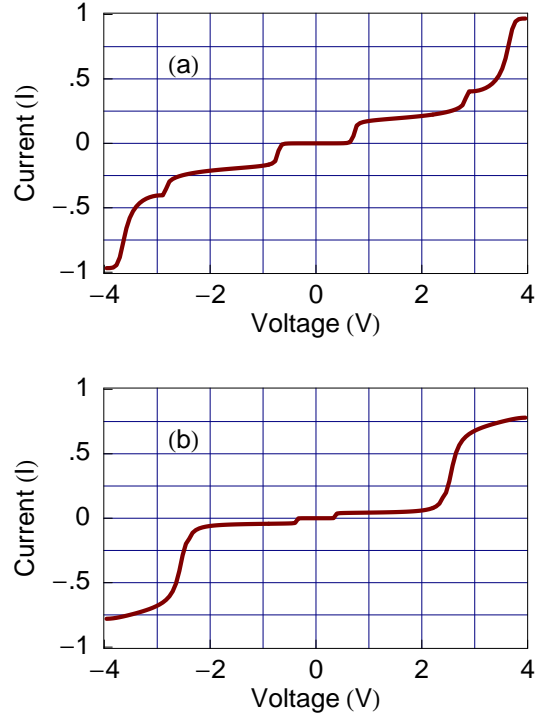}}\par}
\caption{Current $I$ as a function of the bias voltage $V$ for some lattice
ribbons with fixed width $N=3$ and varying lengths where (a) $M=2$ and
(b) $M=3$ (for color illustration, see the web version).}
\label{current2}
\end{figure}
fixed width ($N=2$), the threshold bias voltage ($V_{th}$) of electron 
conduction decreases with the increase of the length of the ribbon. 
This reveals a transformation towards the conducting material. Quite in
the same fashion, to see the variation of the threshold bias voltage
$V_{th}$ for other system sizes, in Fig.~\ref{current2} we plot the results
for some nanoribbons with fixed width $N=3$ and varying lengths where (a)
and (b) correspond to the ribbons with $M=2$ and $3$ respectively. The 
results show that the threshold bias voltages decrease much more compared 
to the nanoribbons of width $N=2$. Thus both from Figs.~\ref{current1} and 
\ref{current2} we clearly observe that $V_{th}$ can be regulated very 
nicely by tuning the size (both length and width) of the nanoribbon.
For quite larger ribbons the threshold bias voltage eventually reduces to
zero. This clearly manifests the transformation from the semiconducting
to the conducting material.

\section{Concluding remarks}

To summarize, we have addressed electron transport in honeycomb lattice 
ribbons with armchair edges attached to two semi-infinite one-dimensional 
metallic electrodes within the tight-binding framework. Our results have 
predicted that for ribbons with smaller lengths and widths, a central gap 
in the 
conductance spectrum appears across the energy $E=0$. While, this gap 
decreases gradually as we increase the size of the ribbon and eventually 
it almost vanishes. This reveals a transformation from the semiconducting 
to the conducting behavior, and it has been much more clearly described 
from the presented current-voltage characteristics which provide that the 
threshold bias voltage of electron conduction decreases gradually with the 
increase of the size of the ribbon. 

This is our first step to describe how the electron transport in honeycomb
lattice ribbons with armchair edges depends on the length and width of the
ribbons. Here we have made several realistic assumptions by ignoring the 
effects of the electron-electron correlation, disorder, temperature, etc. 
All these effects can be incorporated quite easily with this present 
formalism and we need further study in such systems. In our next work we 
are investigating the electron transport properties in honeycomb lattice 
ribbons with zigzag edges.

\end{document}